\documentclass[12pt]{article}
\usepackage{epstopdf}
\usepackage{epsfig}
\usepackage{graphicx}
\usepackage{a4}
\usepackage{amsmath}
\usepackage{latexsym}
\usepackage{cite}

\usepackage{lineno}
\usepackage{color}
\usepackage{colordvi}

\graphicspath{{Pics/}}
\DeclareGraphicsExtensions{.eps,.ps}

\usepackage{pslatex}
\usepackage[latin1]{inputenc}
\usepackage[T1]{fontenc}

\newcommand{\msbar}{$\overline{\text{MS}}\, $}

\begin{document}

\begin{titlepage}
\noindent
DESY 12-233 \hfill December 2012\\
DO-TH 12/35 \\
LPN 12-132 \\
SFB/CPP-12-96 \\
\vspace{1.3cm}

\begin{center}
  {\bf 
\Large
Precise charm-quark mass from deep-inelastic scattering
  }
  \vspace{1.5cm}

  {\large
    S. Alekhin$^{\,a,b}$, 
    J. Bl\"umlein$^{\,b}$, 
    K. Daum$^{\,c,}$\footnote{Permanent address: DESY, 
    Notkestra{\ss}e 85, D--22607 Hamburg, Germany }, 
    K. Lipka$^{\,d}$
    and
    S. Moch$^{\,b,e}$
  }\\
  \vspace{1.2cm}

  {\it 
    $^a$Institute for High Energy Physics \\
    142281 Protvino, Moscow region, Russia\\
    \vspace{0.2cm}
    $^b$Deutsches Elektronensynchrotron DESY \\
    Platanenallee 6, D--15738 Zeuthen, Germany \\
    \vspace{0.2cm}
    $^c$Bergische Universit{\"a}t Wuppertal \\ 
    Gau{\ss}stra{\ss}e 20, D-42097 Wuppertal, Germany \\
    \vspace{0.2cm}
    $^d$Deutsches Elektronensynchrotron DESY \\
    Notkestra{\ss}e 85, D--22607 Hamburg, Germany \\
    \vspace{0.2cm}
    $^e$ II. Institut f\"ur Theoretische Physik, Universit\"at Hamburg \\
    Luruper Chaussee 149, D-22761 Hamburg, Germany 
  }
  \vspace{2.4cm}

\large
{\bf Abstract}
\vspace{-0.2cm}
\end{center}
We present a determination of the charm-quark mass in the \msbar\ scheme 
using the data combination of charm production cross section measurements in deep-inelastic scattering at HERA. 
The framework of global analyses of the proton structure accounts for 
all correlations of the charm-quark mass with the other non-perturbative
parameters, most importantly the gluon distribution function in the proton and the strong coupling constant $\alpha_s(M_Z)$.
We obtain at next-to-leading order in QCD the value 
$m_c(m_c) = 1.15\, \pm 0.04 (\text{exp})\,^{+0.04}_{-0.00} (\text{scale})$ GeV 
and 
at approximate next-to-next-to-leading order 
$m_c(m_c) = 1.24\, \pm 0.03 (\text{exp})\,^{+0.03}_{-0.02} (\text{scale})\,^{+0.00}_{-0.07} (\text{theory})$ GeV 
with an accuracy competitive with other methods.
\vfill
\end{titlepage}

%
% ----------------------------------------------------------------------------------------------
%
\newpage

\section{Introduction}
The charm quark is the lightest of the heavy quarks. 
Yet, the value of its mass $m_c$ is much larger than the scale $\Lambda_{\rm QCD}$ 
of Quantum Chromodynamics (QCD), i.e., $m_c \gg \Lambda_{\rm QCD}$. 
Thus, scattering processes involving charm quarks are subject to QCD dynamics 
at scales of the order of $m_c$, where perturbative QCD predictions apply.
This offers the opportunity to extract $m_c$ by comparing experimental data 
for an appropriate observable to quark mass dependent theoretical predictions in perturbative QCD. 
This procedure does require some care, though.
After all, quark masses are formal parameters of the QCD Lagrangian, but do not belong to the set
of observables in Quantum Field Theory. Quarks and gluons do not belong to the asymptotic states 
at $t \rightarrow \pm \infty$ and already on grounds of the LSZ-theorem their mass is not the usual
mass of a stable elementary particle, like the electron. No free quarks are observed in nature. As
known from pertubation theory, the QCD corrections to the quark masses are renormalization scheme-dependent.
Any quantification of these formal parameters necessarily assumes a definite choice of scheme a priori.  

In the past, high precision cross-section data from $e^+e^-$-collisions have been the basis for 
such charm quark mass determinations.
The available data from $e^+e^-$-annihilation into hadrons span a large range of center-of-mass energies 
and can be used in QCD sum rule analyses based on perturbative QCD predictions to high orders in the
coupling constant 
resulting in precise $m_c$ values from scattering processes with time-like kinematics, 
see, e.g.,~\cite{Beringer:2012}.
The recently available high precision data for charm quark production in
deep-inelastic scattering (DIS) at the HERA collider 
now provide the attractive opportunity for a $m_c$ extraction from scattering processes with space-like kinematics.
This is interesting per se for consistency tests of the Standard Model. 
Moreover, the precision now reached by the DIS measurements allows for an $m_c$ 
determination with an accuracy comparable to the one achieved in QCD sum rule analyses.

In the present paper we use the new data combination of charm production cross section 
measurements in DIS at HERA~\cite{Abramowicz:1900rp} to determine $m_c$ in the \msbar\ scheme 
by comparing to QCD predictions at next-to-leading (NLO) and next-to-next-to-leading (NNLO) order.
We apply the formalism developed in Ref.~\cite{Alekhin:2010sv} and fit $m_c$ to the cross section data 
together with all other non-perturbative parameters, 
of which the gluon distribution function in the proton and the strong coupling constant $\alpha_s(M_Z)$
in particular exhibit a significant correlation with $m_c$.
For this purpose, we update the parton distribution function (PDF) analysis ABM11~\cite{Alekhin:2012ig} 
with the new combined HERA data~\cite{Abramowicz:1900rp} 
included.
Like ABM11, also the new variant of the present paper 
uses the \msbar renormalization scheme for $\alpha_s(M_Z)$ and the heavy-quark masses.
It is performed in the so-called fixed-flavor number (FFN) scheme for $n_f=3$ light quarks to be
dealt with massless.
The latter feature is rather important because in a global fit such as ABM11 
already the data for completely inclusive DIS measurements from HERA put
significant constraints on the value of $m_c$ due to the correlations mentioned.
The FFN scheme allows for a well-defined description of open charm production in QCD, 
and the radiative corrections, i.e., the Wilson coefficients of the 
hard scattering process, are available exactly to NLO~\cite{Laenen:1992zk,Riemersma:1994hv,Bierenbaum:2009zt}
(see also Ref.~\cite{Harris:1995tu}) and to NNLO in an approximation 
for the most important parts, 
that is the gluon and quark pure-singlet Wilson coefficients~\cite{Kawamura:2012cr}.

The present study complements a previous determination of the $c$-quark mass~\cite{Alekhin:2012un} 
in the \msbar\ scheme based on data from the H1 collaboration~\cite{Aaron:2009jy,Aaron:2009af} 
for open charm production. 
Those data are available in differential distributions 
so that the effect of value of $m_c$ on the extrapolation to the unmeasured region for 
the inclusive cross section has been carefully examined.
In this way, Ref.~\cite{Alekhin:2012un} has obtained the \msbar\ mass $m_c(\mu_r=m_c) \equiv m_c(m_c)$ 
for the renormalization scale choice $\mu_r=m_c$ at NLO to 
$m_c(m_c) = 1.27\pm 0.05 (\text{exp})^{+0.06}_{-0.01}(\text{scale})$ GeV 
and at approximate NNLO to 
$m_c(m_c) = 1.36\pm 0.04 (\text{exp})^{+0.04}_{-0.00}(\text{scale})\pm 0.1 (\text{theory})$ GeV, 
respectively.

The present paper is organized as follows. 
In Sec.~\ref{sec:data} we briefly recount the essential features of the data combination of Ref.~\cite{Abramowicz:1900rp}.
Sec.~\ref{sec:analysis} contains the analysis and the new result for $m_c(m_c)$ 
together with a detailed discussion on the impact of
the new data set on the fit and the correlations of $m_c$ with the gluon
distribution and the strong coupling $\alpha_s(M_Z)$.
We conclude in Sec.~\ref{sec:concl} emphasizing that the accuracy of the $m_c$
determination from DIS data becomes competitive with other methods, e.g., QCD sum rule analyses.

\section{Data}
\label{sec:data}
The $c$-quark mass determination is conducted within the framework of a global analyses 
provided by the ABM11 fit~\cite{Alekhin:2012ig}.
The ABM11 analyses~\cite{Alekhin:2012ig} has evolved from the previous ABKM09 fit~\cite{Alekhin:2009ni} 
and is based on world data for deep-inelastic scattering from
HERA, and fixed target experiments and the Tevatron results on the Drell-Yan process.
These data are supplemented by the recently published 
combined charm production cross sections in DIS at HERA~\cite{Abramowicz:1900rp}
and are used as input for the QCD analysis at NLO and NNLO. 

Reduced cross sections for charm production were measured in the kinematic
range of photon virtuality $2.5 \le Q^2 \le 2000\, {\rm GeV}^2$ and Bjorken scaling
variable $3 \cdot 10^{-5} \le x \le 5 \cdot 10^{-2}$. 
The measurement was based on the combination of results obtained by using different charm 
tagging techniques: the reconstruction of $D$ or $D^*$ mesons, the
identification of muons from semi-leptonic decays of charmed
hadrons or by exploiting the long lifetime in charmed hadron decays. 
The individual measurements were performed in different experimentally
accessible (visible) phase space regions, depending on the experimental
technique applied or on the different acceptances of the detector components
used. For $D$-meson and muon production, the visible cross section
measurements were extrapolated to the full phase space using predictions from
perturbative QCD to NLO in the FFN scheme~\cite{Laenen:1992zk,Riemersma:1994hv,Harris:1995tu}.
The quoted uncertainties in the extrapolation include those due to
the variations of the factorization and renormalization scales, $\mu_f$, $\mu_r$,
simultaneously by a factor of $1/2$ and $2$ around the nominal scale,   
as well as of the charm quark mass in the range $1.35< m_c^{\rm pole}< 1.65\, {\rm GeV}$
for the pole mass $m_c^{\rm pole}$ used in Refs.~\cite{Laenen:1992zk,Riemersma:1994hv,Harris:1995tu}.

The correlated systematic uncertainties and the normalization of the different
measurements were accounted for in the combination procedure such that one
consistent data set has been obtained. Since different experimental techniques
of charm tagging were employed, the combination led to a significant reduction
of the statistical and systematic uncertainties. However, due the combination
procedure the information about the extrapolation factors and their
uncertainties for the individual input data sets cannot be provided.
Therefore, a detailed analysis similar to the previous one performed in Ref.~\cite{Alekhin:2012un} taking
into account the dependence of the extrapolation factor on the assumption of
the charm mass in the underlying theory is not possible here. 
Instead, the results on the combined reduced charm cross sections at
particular kinematical points ($x, Q^2$) are used in the current analysis with
account of the correlations of the uncertainties as provided by the 
experiments~\cite{h1zeuscombo:2012}.

\section{Analysis}
\label{sec:analysis}
The theoretical framework applied in the present analysis of the combined HERA data~\cite{Abramowicz:1900rp} 
essentially coincides with the one used earlier in the determination of the $c$-quark mass~\cite{Alekhin:2012un} 
with the H1 data  on open charm production~\cite{Aaron:2009jy,Aaron:2009af}. 
We compute the heavy-quark contribution to the DIS cross section in the scheme 
with $n_f=3$ massless flavors in the initial state. 
The running-mass definition is employed for the heavy-quark Wilson coefficients, which comprise 
the NLO terms~\cite{Alekhin:2010sv} derived from the calculations performed
with the pole mass definition~\cite{Laenen:1992zk} and the NNLO terms~\cite{Kawamura:2012cr}. 
The latter are denoted by NNLO$_\text{approx}$ in the following, 
obtained by interpolation between existing soft-gluon threshold resummation results and 
approximate relations for the Wilson coefficients at $Q^2\gg m^2$ 
taking advantage of selected Mellin moments for the massive operator-matrix elements at NNLO 
given in 
Refs.~\cite{Buza:1995ie,Bierenbaum:2007qe,Bierenbaum:2008yu,Bierenbaum:2009mv,Ablinger:2010ty}. 
and the massless 3-loop Wilson coefficients \cite{Vermaseren:2005qc}.
The residual interpolation uncertainty which appears due to the finite number of Mellin moments
being known 
is quantified by two options, $c_{2}^{\,(2),A}$ and $c_{2}^{\,(2),B}$,  
for the constant terms in the Wilson coefficients at NNLO~\cite{Kawamura:2012cr}.
In the present analysis the shape of the NNLO correction is defined as 
a linear interpolation between these options using the ansatz
\begin{equation}
\label{eq:inter}
  c_{2}^{\,(2)} \,=\, 
  (1-d_N) c_{2}^{\,(2),A} + d_N c_{2}^{\,(2),B} 
  \, .
\end{equation}
The {\tt Fortran} code {\tt OPENQCDRAD} for the numerical computation of all cross sections 
in the present analysis is publicly available~\cite{openqcdrad:2012}.

Our determination of $m_c$ is based on the 3-flavor
ABM11 PDFs~\cite{Alekhin:2012ig}.
However, those PDFs were obtained at the fixed value of 
$m_c(m_c)=1.27~{\rm GeV}$. In order to provide a consistent treatment of the 
PDF dependence on $m_c$ we employ in the present analysis a set of $m_c$-dependent PDFs
produced by interpolating between the variants of the ABM11 fit with the value
of $m_c(m_c)$ scanned over the range of $0.9-1.35$~GeV. 
By fitting to the combined HERA charm data in this way we obtain the following 
$c$-quark mass values in the \msbar\ scheme
\begin{eqnarray}
  \label{eq:mcabm11-nlo}
  m_c(m_c) \,\,=&
  1.20\, \pm 0.05 (\text{exp})
  \hspace*{30mm}
  &{\rm NLO}
  \, , 
  \\
  \label{eq:mcabm11-nnlo}
  m_c(m_c) \,\,=&
1.30\, \pm 0.04 (\text{exp})
  \hspace*{30mm}
  &{\rm NNLO_\text{approx}}
  \, .
\end{eqnarray}
Here the NNLO value corresponds to $d_N=-0.4$ which provides the   
best agreement with the data
in line with the approach of Ref.~\cite{Alekhin:2012un}.
The experimental uncertainties in $m_c(m_c)$ are calculated by propagation of the 
errors in the data, taking into account the systematic error correlations.
For the combined HERA data~\cite{Abramowicz:1900rp} they stem from 48 sources 
including the extrapolation of the visible charm production cross 
section to the full phase space\footnote{The combined HERA data on  
open charm production with their systematic uncertainties used 
in the present analysis are available from {\tt http://arxiv.org}
as an attachment to the arXiv version of the present paper.}. 
This extrapolation is sensitive to 
the calculation details, such as fragmentation-model parameters, the PDFs, the value of $m_c$, etc. 
The corresponding systematic errors encode the impact 
of the sensible variation of these parameters on the cross section values. 
Ideally, the extrapolation correction
has to be calculated in the analysis iteratively, in parallel with fitting $m_c$ and the PDFs, as it has been 
done in the earlier determination of $m_c$ in Ref.~\cite{Alekhin:2012un} based on the selected set of 
the H1 open charm production data. 
As discussed in Sec.~\ref{sec:data}, this approach is inapplicable in the present analysis 
because the necessary information about the visible phase 
space is lost in the combination of the H1 and ZEUS data. 
The extracted value of $m_c$ thus faces a procedural bias
due to the fact that the extrapolation corrections are calculated for a fixed value of $m_c$. 
However, the corresponding uncertainty was estimated in Ref.~\cite{Abramowicz:1900rp} 
by a conservative variation of the input used in the extrapolation correction. 
Therefore, the quoted experimental uncertainties in $m_c$ must exceed this bias. 

The central values of $m_c$ in Eqs.~(\ref{eq:mcabm11-nlo}) and (\ref{eq:mcabm11-nnlo}) are lower 
than those in our earlier determination in Ref.~\cite{Alekhin:2012un}. 
In particular, this difference can be explained by a 
shift of the data obtained by the H1 and ZEUS experiments 
in the process of their combination, cf. Ref.~\cite{Abramowicz:1900rp} for details. 
Besides, the NNLO correction employed in Ref.~\cite{Alekhin:2012un} corresponds to the 
interpolation parameter $d_N=-0.6$, which is somewhat different from 
the one obtained in the present analysis, which causes an additional 
shift of $m_c(m_c)$ in NNLO. However, in any case the values of $m_c(m_c)$
in Eqs.~(\ref{eq:mcabm11-nlo}) and (\ref{eq:mcabm11-nnlo}) are compatible with the 
results of Ref.~\cite{Alekhin:2012un} within the uncertainties.

To study the sensitivity of the $m_c$ determination to the particular choice
of PDFs we repeat our analysis considering other 3-flavor PDFs.
For this purpose, we take in all cases the nominal PDFs obtained with the fixed values of the $c$-quark mass.
The NNLO values of $m_c$ obtained in this way demonstrate good agreement, cf. Tab.~\ref{tab:comp}. 
At NLO only the ABM11~\cite{Alekhin:2012ig} and GJR~\cite{Gluck:2007ck,JimenezDelgado:2008hf} 
results coincide, while lower values are obtained in case of
MSTW08~\cite{Martin:2009iq} and NN21~\cite{Ball:2011mu}.
This difference may partly appear due to a spread in the $c$-quark mass taken in different PDF fits. 
However, the difference between the ABM11 results obtained with and without 
taking into account the $m_c$-dependence of PDFs is of ${\cal O}(10)~{\rm MeV}$, 
cf. Tab.~\ref{tab:comp} and Eqs.~(\ref{eq:mcabm11-nlo}) and (\ref{eq:mcabm11-nnlo}). 
This may point to other reasons for this difference. 
In fact, it is also correlated with the scheme used in the PDF fits. 
While the ABM11 and JR PDFs are based on the 3-flavor scheme,
the MSTW and NNPDF analysis are performed with different versions of a 
general-mass variable-flavor-number (GMVFN) scheme. 
In particular, this explains the difference at NLO between the MSTW and ABM/GJR results since
the GMVFN scheme commonly deviates at NLO from the 3-flavor one to a larger extent than at NNLO. 
Recall also, that all PDF fits except ABM11, refer to the on-shell scheme for heavy quarks 
and compare to theoretical predictions using the pole mass $m_c^{\rm pole}$. 

\begin{table}[h]
\begin{center}
\begin{tabular}{|c|c|c|c|c|} \hline
  & ABM11~\cite{Alekhin:2012ig}   &   JR(GJR)~\cite{Gluck:2007ck,JimenezDelgado:2008hf}    &  MSTW08~\cite{Martin:2009iq}   &   NN21~\cite{Ball:2011mu} \\ \hline
NLO  &  1.21  &   1.21  &   1.12   &  1.01 \\ \hline 
NNLO &  1.28  &   1.27 &  1.29  &  -- \\ \hline
\end{tabular}
\end{center}
\caption{
  \label{tab:comp}
  \small The value of $m_c(m_c)$ in GeV obtained from the 
analysis of the combined HERA data on open charm 
production~\cite{Abramowicz:1900rp} with different 3-flavor PDFs in NLO and 
NNLO. Note, the ABM11 values are different from the ones in 
Eqs.~(\ref{eq:mcabm11-nlo}) and (\ref{eq:mcabm11-nnlo}) since the latter were obtained within the 
$m_c$-dependent variant of the ABM11 PDFs.}
\end{table}

\begin{figure}[hhh]
\center            
  \includegraphics[width=0.9\textwidth]{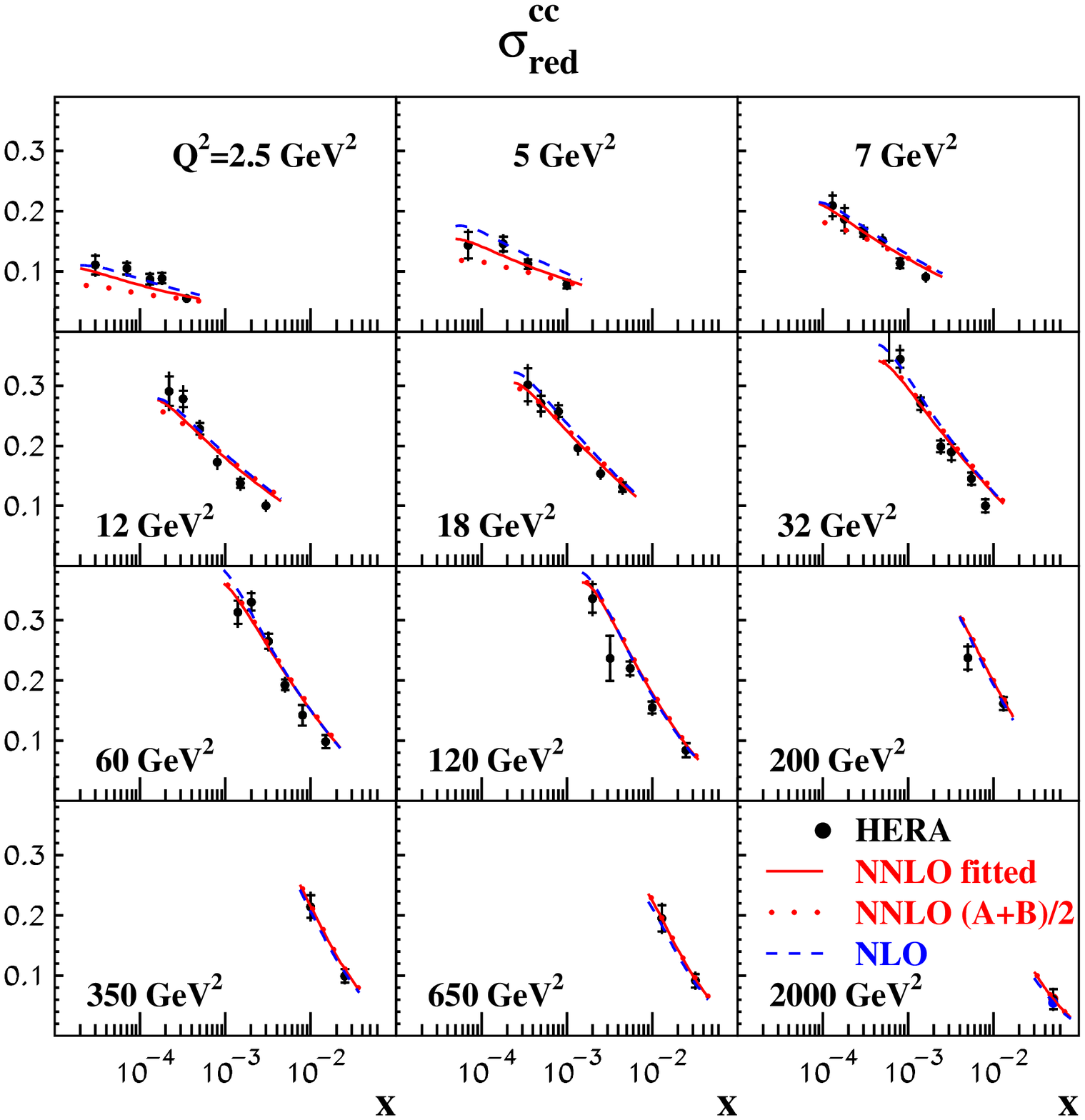}
\setlength{\unitlength}{1cm}                               
\caption{\label{fig:hera}                                                                    
  The combined HERA data on the reduced cross section for the open 
  charm production~\cite{Abramowicz:1900rp} versus $x$ at different
  values 
  of $Q^2$ in comparison with the result of the present analysis at   
  NLO (dashed line) and NNLO (solid line).
  A variant of the fit based on the option (A+B)/2 of the NNLO Wilson
  coefficients of Ref.~\cite{Kawamura:2012cr}, 
  cf. Eq.~(\ref{eq:inter}), is displayed for comparison (dotted line).
} 
\end{figure}

Although Eqs.~(\ref{eq:mcabm11-nlo}) and (\ref{eq:mcabm11-nnlo}) for $m_c(m_c)$ 
are based on a consistent treatment of the PDF's $c$-quark mass dependence, 
the constraints on the variation of those PDFs with $m_c$ 
imposed by the data included in the ABM11 fit are not yet taken into account
in the the determination of those numbers.
To take advantage of the sensitivity of charm production in neutrino-nucleon scattering \cite{Bazarko:1994tt,Goncharov:2001qe}
and the inclusive DIS to the charm mass we also perform the NLO and NNLO variants of ABM11 fit, 
which includes those data together with the HERA charm data of Ref.~\cite{Abramowicz:1900rp} added
and the value of $m_c(m_c)$ considered as a fitted 
parameter\footnote{To allow for a variation of the
factorization scale in the present 
analysis a cut of $Q^2>2~{\rm GeV}^2$ is imposed 
on the data for dimuon production in neutrino-nucleon 
DIS~\cite{Bazarko:1994tt,Goncharov:2001qe}, while in the 
analysis of Ref.~\cite{Alekhin:2012ig} these data with      
$Q^2\simeq 1~{\rm GeV}^2$ were used.}. 
From these versions of the fit we obtain the values of 
\begin{eqnarray}
  \label{eq:mcres-nlo}
  m_c(m_c) \,\,=&
  1.15\, \pm 0.04 (\text{exp})\,^{+0.04}_{-0.00} (\text{scale})
  \hspace*{30mm}
  &{\rm NLO}
  \, ,
\end{eqnarray}
\begin{eqnarray}
%  \\
  \label{eq:mcres-nnlo}
  m_c(m_c) \,\,=&
  1.24\, \pm 0.03 (\text{exp})\,^{+0.03}_{-0.02} (\text{scale})\,^{+0.00}_{-0.07} (\text{th}),
  \hspace*{14mm}
  &{\rm NNLO_\text{approx}}
  \, ,
\end{eqnarray}
where the NNLO value corresponds to $d_N=-0.1$. This  
provides the best description of data with $\chi^2$ normalized by the number
of data points ($NDP$), 
$\chi^2/NDP = 3459/3080$ for the whole data set and 
$\chi^2/NDP = 61/52$ for the combined HERA charm data, cf. also Fig.~\ref{fig:hera}.
At the same time, the option B of the massive NNLO correction of 
Ref.~\cite{Kawamura:2012cr} corresponding to $d_N=1$
is clearly disfavored by the data giving 
$\chi^2/NDP=115/52$ for the HERA charm data and $\chi^2/NDP=3547/3080$ for the whole data set. 
Therefore we estimate the uncertainty due to the massive NNLO correction 
choice as variation between the values of $m_c(m_c)$ obtained with $d_N=-0.1$ and $d_N=0.5$ in Eq.~(\ref{eq:inter}).
This yields the value of $0.07~{\rm GeV}$ quoted in 
Eq.~(\ref{eq:mcres-nnlo}) as an estimate of the theoretical uncertainty. 
The scale uncertainty in $m_c(m_c)$ is calculated as a variation 
due to a change in the factorization scale by a factor of $1/2$ and $2$ around the nominal value of $\sqrt{m_c^2+\kappa Q^2}$, 
where $\kappa=4$ for neutral-current and $\kappa=1$ for charge-current heavy-quark production, respectively. 
For the NLO case both these variations lead to 
an increase in $m_c(m_c)$ and we select the bigger shift as the 
uncertainty due to the scale variation.
The NNLO scale uncertainty in $m_c(m_c)$ is asymmetric and smaller than 
the NLO one, in line with the estimates of Ref.~\cite{Kawamura:2012cr}.
The experimental error of $m_c$ is reduced due to the constraints on the PDFs 
by the inclusive DIS data. 
The theoretical error due to missing higher order corrections is the 
dominant source of uncertainty in $m_c$. 

The central value of $m_c(m_c)$ obtained at NLO in Ref.~\cite{Abramowicz:1900rp} 
for the combined HERA data including the data  on charm production, i.e.,  
$m_c(m_c) = 1.26\, \pm 0.05 (\text{exp})$~GeV, 
turns out to be bigger than our NLO result. It is important to 
note that this value is obtained from a scan of
$m_c(m_c)$ and not in a simultaneous fit of the PDFs and the charm quark mass.
Also, partially the difference to our result can be explained by the different cuts on $Q^2$ imposed in the analysis of Ref.~\cite{Abramowicz:1900rp} and ours. 
By changing our cut of $Q^2>2.5~\text{GeV}^2$ to the cut of 
$Q^2>3.5~\text{GeV}^2$ used in~\cite{Abramowicz:1900rp} 
we get a shift of $+0.03~\text{GeV}$ our NLO value of $m_c(m_c)$ in Eq.~(\ref{eq:mcres-nlo}).
Another source for the difference is the data on dimuon production 
in neutrino-nucleon DIS~\cite{Bazarko:1994tt,Goncharov:2001qe} included in ABM11.
By excluding this data set, we obtain a shift of $+0.04~\text{GeV}$ for 
the $c$-quark mass in Eq.~(\ref{eq:mcres-nlo}).
Note, that the value of $m_c(m_c)$ of Ref.~\cite{Abramowicz:1900rp} is also systematically 
bigger than the NLO entries in Tab.~\ref{tab:comp} which are obtained with fixed PDFs. 
Therefore the remaining difference between Eq.~(\ref{eq:mcres-nlo}) and $m_c(m_c)$ of Ref.~\cite{Abramowicz:1900rp} 
is evidently also related to particularities of the shape of HERA PDFs 
used in Ref.~\cite{Abramowicz:1900rp}.

\begin{figure}[hhh]
\center            
  \includegraphics[width=0.9\textwidth]{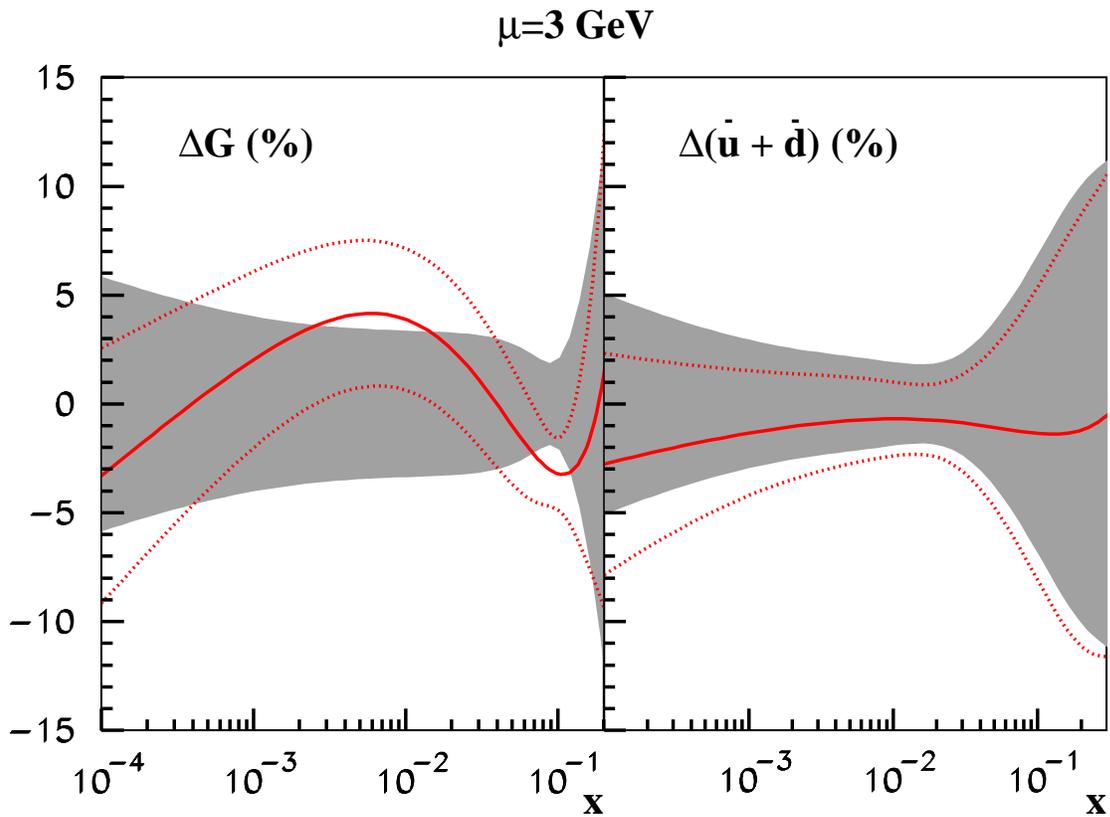}
\setlength{\unitlength}{1cm}                               
\caption{\label{fig:pdfs}
  The relative change in the NNLO gluon (left) and non-strange sea 
  (right) distributions obtained in the present analysis
  with respect to the ABM11 PDFs (solid lines). The relative uncertainties 
  in the PDFs are displayed for comparison (shaded area: ABM11, dotted lines: 
  present analysis).
}
\end{figure}

Let us finally discuss a number of cross checks. 
Operating in the framework of a global analysis of the proton structure as provided by ABM11 
offers the possibility to account consistently for all correlations of the
$c$-quark mass with non-perturbative parameters of the fit of which 
the gluon distribution function and the strong coupling constant $\alpha_s(M_Z)$ 
exhibit the strongest correlation with $m_c$.
We observe, that the shape of the gluon distribution obtained in the present fit is 
somewhat modified with respect to the ABM11 PDFs, cf. Fig.~\ref{fig:pdfs}.
However, the changes are basically found to be within the PDFs uncertainties. 
The sea distribution is affected to a lesser extent and the other PDFs are practically unchanged. 

\begin{figure}[hhh]
\center            
  \includegraphics[width=0.9\textwidth]{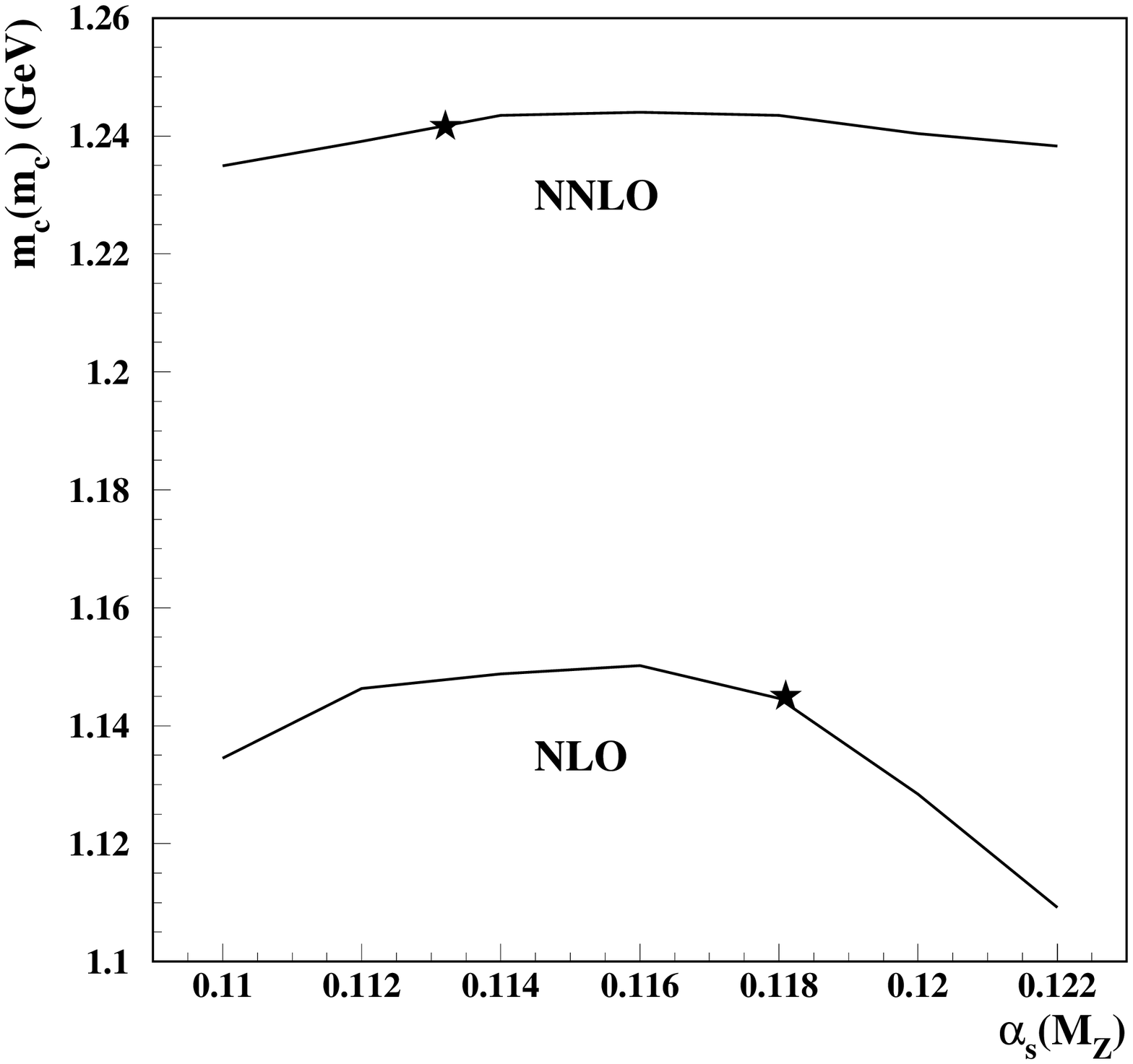}
\setlength{\unitlength}{1cm} 
\caption{\label{fig:alpha}
  The values of $m_c(m_c)$ obtained in the NLO and NNLO variants of
  the ABM11 fit with the combined HERA charm data~\cite{Abramowicz:1900rp} 
  included and the value of $\alpha_s(M_Z)$ fixed. 
  The position of the star displays the result with the value of $\alpha_s(M_Z)$ fitted~\cite{Alekhin:2012ig}.
}
\end{figure}

The correlation of the fitted value of $m_c$ with the 
strong coupling constant $\alpha_s(M_Z)$ is shown in Fig.~\ref{fig:alpha} 
for a variation of the value of $\alpha_s(M_Z)$ in the range $\alpha_s(M_Z)=0.110 - 0.122$. 
Recall, that the analysis of ABM11~\cite{Alekhin:2012ig} has obtained 
$\alpha_s(M_Z) = 0.1180 \pm 0.0012$ at NLO and 
$\alpha_s(M_Z) = 0.1134 \pm 0.0011$ at NNLO as best fits.
Fig.~\ref{fig:alpha} demonstrates a remarkable stability of the $c$-quark mass both at NLO and NNLO. 
Considering a variation of $0.115 \le \alpha_s(M_Z) \le 0.119$ 
the shift of $\Delta m_c(m_c)$ is confined within an interval of 20~MeV 
for the NLO case and 
for a range of $0.110 \le \alpha_s(M_Z) \le 0.114$ at NNLO 
within an interval of 10~MeV only. 
This is to be compared with the $\alpha_s(M_Z)$ dependence inherent in QCD sum rule analyses. 
For example, for a variation of $0.113 \le \alpha_s(M_Z) \le 0.119$ 
Ref.~\cite{Dehnadi:2011gc} observes a linear growth of the value of $m_c(m_c)$
with a maximal shift of $\Delta m_c(m_c) = 25~{\rm MeV}$ (cf. Fig.~11a in~\cite{Dehnadi:2011gc}).
In contrast, the numbers for $m_c(m_c)$ determined in Eqs.~(\ref{eq:mcres-nlo}) and (\ref{eq:mcres-nnlo}) 
do not carry such bias with respect to the value of the strong coupling constant.

To conclude the discussion we also convert the values of $m_c(m_c)$ in Eqs.~(\ref{eq:mcres-nlo}) and (\ref{eq:mcres-nnlo}) 
to the on-shell scheme. Using the well-known relations for the scheme
transformation as encoded in~\cite{Chetyrkin:2000yt} and the values for $\alpha_s(M_Z)$ of ABM11 at NLO and NNLO, 
we obtain 
\begin{eqnarray}
  \label{eq:mcpole-nlo}
  m_c^{\rm pole} \,\,=&
  1.35\, \pm 0.05 (\text{exp})\,^{+0.05}_{-0.00} (\text{scale})
  \hspace*{30mm}
  &{\rm NLO}
  \, ,
  \\
  \label{eq:mcpole-nnlo}
  m_c^{\rm pole} \,\,=&
  1.59\, \pm 0.04 (\text{exp})\,^{+0.04}_{-0.03} (\text{scale})\,^{+0.00}_{-0.09} (\text{th}),
  \hspace*{14mm}
  &{\rm NNLO_\text{approx}}
  \, .
\end{eqnarray}
As to be expected, the numerical values for $m_c^{\rm pole}$ are larger than the 
values given in Eqs.~(\ref{eq:mcres-nlo}) and (\ref{eq:mcres-nnlo})
and those positive corrections grow in size, i.e., the shift of the central value amount to 
$\Delta m_c(m_c) = 200~{\rm MeV}$ at NLO and $\Delta m_c(m_c) = 350~{\rm MeV}$ at NNLO.
The increasing spread between the numbers in Eqs.~(\ref{eq:mcpole-nlo}) and (\ref{eq:mcpole-nnlo}) 
can illustrate the poor perturbative convergence of the pole mass scheme which
is particularly pronounced at the low scales relevant for DIS charm production.

\section{Conclusions}
\label{sec:concl}

The new combined HERA data for charm production cross section measurements 
in DIS allows for a precise determination of the charm-quark mass 
in the \msbar\ scheme by comparing to QCD theory predictions in the 
FFN scheme at NLO and NNLO.
Embedding the data analysis in a global fit takes advantage of a 
well-established theory framework and, simultaneously accounts for all 
correlations with other non-perturbative parameters, of which the gluon PDF in
the proton and the strong coupling constant $\alpha_s(M_Z)$ are most important and
have been studied in detail.

The effect of the HERA DIS charm data on the extraction of $m_c(m_c)$ 
has been demonstrated in Eqs.~(\ref{eq:mcabm11-nlo}), (\ref{eq:mcabm11-nnlo}).
Yet, the full potential for a precision determination of $m_c(m_c)$ unfolds 
in a global fit due the additional constraints imposed by the inclusive HERA data 
and those from neutrino-nucleon DIS.
Thus, the best values for the $c$-quark mass are 
$m_c(m_c) = 1.15\, \pm 0.04 (\text{exp})\,^{+0.04}_{-0.00} (\text{scale})$ GeV 
at NLO and 
$m_c(m_c) = 1.24\, \pm 0.03 (\text{exp})\,^{+0.03}_{-0.02} (\text{scale})\,^{+0.00}_{-0.07} (\text{theory})$ GeV 
at approximate NNLO, cf. Eqs.~(\ref{eq:mcres-nlo}) and (\ref{eq:mcres-nnlo}), 
although the accuracy of the latter determination still suffers from 
missing information on the three-loop Wilson coefficients for neutral current 
DIS heavy quark production at small-$x$ and small values of $Q^2$. 
This implies an additional theoretical uncertainty on $m_c(m_c)$ estimated to be 
in the range $- 70 \le \Delta m_c \le 0$ MeV.
The obtained values in Eqs.~(\ref{eq:mcres-nlo}) and (\ref{eq:mcres-nnlo}) 
are compatible with the previous analysis of Ref.~\cite{Alekhin:2012un} and 
with the world average $m_c(m_c) = 1.275 \pm 0.025~\text{GeV}$  
as summarized by the particle data group \cite{Beringer:2012}.
The accuracy of the determination is competitive with other approaches, 
e.g., from scattering reactions in time-like kinematics.

\subsection*{Acknowledgments}
We acknowledge fruitful discussions with R.~Pla\v{c}akyt\.{e}.
This work has been supported in part by Helmholtz Gemeinschaft under contract 
VH-HA-101 ({\it Alliance Physics at the Terascale}), 
VH-NG-401 ({\it Young Investigator group "Physics of gluons and heavy quarks"}), 
by the Deutsche Forschungsgemeinschaft in Sonderforschungs\-be\-reich/Trans\-regio~9 and 
by the European Commission through contract PITN-GA-2010-264564 ({\it LHCPhenoNet}).

{\footnotesize
%\bibliography{update.bib}
%\bibliographystyle{h-physrev5.bst}

}

\end{document}